**Title**

Multicenter Assessment of Augmented Reality Registration Methods for Image-guided Interventions

**Authors**

Ningcheng Li [1], Jonathan Wakim [2], Yilun Koethe [1], Timothy Huber [1], Terence Gade [2], Stephen Hunt [2], Brian Park [1]

1. Dotter Department of Interventional Radiology, Oregon Health & Science University School of Medicine, Portland, OR
2. Department of Radiology, Perelman School of Medicine at the University of Pennsylvania, Philadelphia, PA

**Abstract**

Purpose: To evaluate manual and automatic registration times as well as accuracy with augmented reality during alignment of a holographic 3-dimensional (3D) model onto the real-world environment.

Method: 18 participants in various stages of clinical training across two academic centers registered a 3D CT phantom model onto a CT grid using the HoloLens 2 augmented reality headset 3 consecutive times. Registration times and accuracy were compared among different registration methods (hand gesture, Xbox controller, and automatic registration), levels of clinical experience, and consecutive attempts. Registration times were also compared with prior HoloLens 1 data.

Results: Mean aggregate manual registration times were 27.7, 24.3, and 72.8 seconds for one-handed gesture, two-handed gesture, and Xbox controller, respectively; mean automatic registration time was 5.3s (ANOVA $p<0.0001$). No significant difference in registration times was found among attendings, residents and fellows, and medical students ($p>0.05$). Significant improvements in registration times were detected across consecutive attempts using hand gestures ($p<0.01$). Compared with previously reported HoloLens 1 experience, hand gesture registration times were 81.7% faster (p<0.05). Registration accuracies were not significantly different across manual registration methods, measuring at 5.9, 9.5, and 8.6 mm with one-handed gesture, two-handed gesture, and Xbox controller, respectively (p>0.05).

Conclusions: Manual registration times decreased significantly with updated hand gesture maneuvers on HoloLens 2 versus HoloLens 1, approaching the registration times of automatic registration and outperforming Xbox controller mediated registration.

**Introduction**

Augmented reality (AR) technology has advanced significantly in recent years. Its ability to display 3-dimensional (3D) datasets, such as computed tomography (CT) and magnetic resonance images (MRI), in true 3D space rather than on flattened 2D screens has been shown to improve anatomical learning [1,2], 3D spatial interpretation [3], and pre-operative planning [4–9]. The added benefit of overlaying such 3D information directly onto patients or real-world environments for procedural guidance and navigation has further pushed its expansion in clinical practice, leading to pilot implementations intra-procedurally across multiple different specialties [4–7,10–13].

The clinical utility of AR using HoloLens 1 (Microsoft, Redmond, WA), one of the first commercially available optical see-through head-mounted displays, has been demonstrated in recent studies to improve spatial understanding and anatomic definition, decrease procedural duration, and reduce fluoroscopy times and radiation exposure [14–17]. Manual registration methods, such as hand-gesture and gaming controller mediated maneuvers, can be implemented for 3D spatial registration of holographic models onto physical objects [18]. The registration accuracy of spatial information onto the real-world environment has been shown to be near or within one centimeter for surgical applications [19–24]. Furthermore, prior work has demonstrated the use of computer vision recognition of a CT grid for automatic registration, which outperformed manual registration methods [18]. However, widespread adoption of AR technologies, including HoloLens-based platforms, in clinical practice has been relatively slow despite the conferred preclinical benefits. One of the main hindrances is likely due to the capabilities limited to the headset devices that can result in unnatural manual manipulations in

AR space. Additionally, the dependence of computer vision recognizable markers or grids, some of which are non-reusable or non-sterile, may further compound the issue for automatic registration. Natural user adaptation of an AR system, fast and accurate registration of 3D information, and freeform spatial manipulation of 3D models are necessary for widespread AR implementation in clinical practice.

The next-generation HoloLens 2 (Microsoft, Redmond, WA), compared to its predecessor, is a more ergonomic headset with an increased field of view, higher processing power, and articulated gesture tracking with greater degrees of freedom (**Table 1**) [25–30]. The improved gesture tracking capabilities provided by HoloLens 2 allow for more intuitive and natural ways of interacting with 3D models not previously possible [31]. In this study, manual and automatic registration methods were developed for HoloLens 1 and HoloLens 2. Registration times and accuracies for aligning a virtual 3D holographic model onto a CT grid across two tertiary academic hospital centers were recorded and compared. Performance differences between the two generations of HoloLens headsets were also compared.

**Methods**

This study was institutional review board exempt at Oregon Health & Science University and at the University of Pennsylvania. A similar workflow previously deployed on HoloLens 1 was implemented for HoloLens 2 [18]. A 3D holographic model of an abdominal CT phantom (CIRS 071B) with an overlying CT grid (Beekley Medical Grid 117) was segmented using ITK-SNAP and projected using HoloLens 2 [32]. Custom code was developed in Unity (Unity Technologies, San Francisco, CA) and C#/.NET framework for model registration using three manual methods: 1) one-handed gestures, 2) two-handed gestures, and 3) Xbox wireless gaming controller (Microsoft, Redmond, WA) interfaced with HoloLens via Bluetooth (**Figure 1**). Automatic registration was performed with computer vision and automated image detection using Vuforia SDK (PTC Technology, Needham, MA) [14,18]. In contrast to HoloLens 1, HoloLens 2 tracking allows for interactions and manipulations with natural articulated hand gestures (**Table 1**), such as simultaneous translation and rotation. This improved capability eliminates the need to toggle between translational and rotational interactions, as often required on HoloLens 1, allowing for much more intuitive maneuvering of the 3D holographic model with HoloLens 2.

Code was distributed and deployed across two tertiary academic hospital centers. Eighteen participants from two centers in various stages of clinical training (three attendings, three trainees including both residents and fellows, and three medical students at each center) attempted alignment using HoloLens 2 with each registration method three consecutive times. Registration times for aligning the 3D model to the CT grid until visual satisfaction and alignment errors within 1 cm in each of the x-, y-, and z-planes were recorded. Registration times were also compared with those previously studied and reported with HoloLens 1 [18].

A single mixed reality image, with virtual CT grid superimposed on actual CT grid, was captured directly from HoloLens 2 after each registration attempt. Dimensions of the CT grid were used for measurement calibration. Precision was determined based on three repeated measurements of the CT grid dimensions over all obtained mixed reality images. Registration accuracy was measured in the captured 2D plane as the longest distance between one of the CT grid markers and its corresponding phantom grid markers, center to center (**Figure 2**), averaged across three different measurements.

Data was presented as mean ± standard error of the mean (SEM). One-way ANOVAs with Tukey's post-hoc multiple comparison tests, repeated measures ANOVAs, Mann-Whitney tests, and t-tests were performed using Prism (GraphPad, San Diego, CA). Two-tailed p-value smaller than 0.05 was deemed statistically significant.

**Results**

Center 1 demonstrated mean manual alignment times of 22.6 ± 3.5s, 25.2 ± 3.5s, and 77.2 ± 19.6s for one-handed gesture, two-handed gesture, and Xbox controller registration methods, respectively, on HoloLens 2. Center 2 demonstrated similar manual alignment times of 32.7 ± 4.6s, 23.4 ± 2.5s, and 68.5 ± 7.2s, respectively (all two-tailed p>0.05 when compared to Center 1). Center 1 reported mean automatic registration time of 7.2 ± 0.6s versus Center 2 with 3.4 ± 0.3s, representing a 52.8% difference (p<0.001).

Manual and automatic registration times from the two centers were combined for HoloLens 2 (**Figure 3**). One-way ANOVA demonstrated significant differences among the different registration methods (p<0.0001) except for the difference between one-handed and two-handed gestures. Post-hoc multiple comparisons tests showed Xbox controller registration as the slowest (72.8 ±10.2s; p<0.01, p<0.001, and p<0.0001 compared to one-handed gesture (27.7 ±3.1s), two-handed gesture (24.3 ±2.1s), and automatic registration (5.3 ±0.6s), respectively). Automatic registration was the fastest (p<0.0001 compared to the other three manual registration methods).

Further analysis was performed based on level of clinical experience. There were no significant differences in registration times among attendings, residents and fellows, and medical students with any of the registration methods (p>0.05, **Supplemental Figure 1**). Significant improvements in registration times were detected across consecutive attempts using hand gestures (one-handed and two-handed combined, 31.7 ±3.6s, 25.4 ±2.0s, 20.8 ±1.5s over the three consecutive attempts, p<0.01, **Figure 4a**) but not with Xbox controller (78.1 ±11.1s, 75.0 ±12.1s, and 65.4 ±14.4s over the three attempts, p>0.05, **Figure 4b**) or automatic registration methods (6.4 ±0.9s, 4.4 ±0.4s, and 5.1 ±0.8s over the three attempts, p>0.05, **Figure 4c**). In particular, during the third attempt, mean hand gesture registration time was 20.8s with the fastest registration time of 7.4s on HoloLens 2.

The HoloLens 2 dataset was combined from both centers and compared to HoloLens 1. In total, mean aggregate hand gesture registration was faster with HoloLens 2 (26.0 ±1.8s) compared to HoloLens 1 (141.9 ±10.0s, p<0.0001), representing an 81.7% reduction. There was no significant difference with manual registration using an Xbox controller (72.8 ±10.2s with HoloLens 2 versus 53.7 ±6.5s with HoloLens 1, p>0.05) as these interactions remained the same on HoloLens 1 and HoloLens 2. Automatic registration was also similar across HoloLens 1 and HoloLens 2 (5.3 ±0.6s with HoloLens 2 versus 5.8 ±0.7s with HoloLens 1, p>0.05).

The average manual measurement precision over 80 different line segments, each with three individual measurements, was 4.3 pixels. This was calibrated and translated to an average measurement precision of 0.8 mm. Registration accuracy was not significantly different across manual registration methods, with one-handed gesture at 5.9 ±0.4 mm, two-handed gesture at 9.5 ±2.3 mm, and Xbox controller registration at 8.6 ±2.0 mm (p>0.05, **Figure 5a**). Attendings, trainees, and medical students demonstrated similar registration accuracies (6.9 ±0.6 mm, 8.9 ± 1.6 mm, and 7.0 ±0.9 mm, respectively, p>0.05; **Figure 5b**). A trend of improving registration accuracies across the three attempts was suggested but not statistically significant (8.8 ±1.3 mm, 8.0 ±1.0 mm, and 5.3 ±0.4 mm for first, second, and third attempts, respectively, p=0.05; **Figure 5c**).

## Discussion

Clinical adoption of AR technologies including HoloLens-based platforms has been slow, partly due to nonintuitive and cumbersome manual manipulations in AR space. HoloLens 2 offered new possibilities with its improved built-in eye and hand tracking capabilities [31]. In this study, registration of a 3D holographic model onto a CT grid using a HoloLens 2 workflow was assessed at two different academic centers and compared retrospectively to the performance of HoloLens 1.

Significant reductions in hand gesture mediated registration times were demonstrated with HoloLens 2 compared to those of HoloLens 1, likely due to a more natural adaptation to the registration maneuvers as comprehensive hand and wrist movements, including pinching and grabbing, could be used for model manipulation. Moreover, freeform spatial manipulation allowed seamless integration of translational and rotational movements, eliminating the need to toggle between modes of maneuver [18]. As a result, mean aggregate hand gesture manual registration time decreased by more than 80% and was significantly shorter than Xbox controller mediated registration. Users also demonstrated significant improvements over the three consecutive attempts. For several participants, hand gesture mediated registration times during the third attempt were less than 10 seconds and on par with the automatic registration times. These results carry significant implications for clinical integration with HoloLens 2 as image-guided interventionalists could benefit from its more intuitive adaptation, ease of use and improvement over time, and ability for rapid manual adjustments in the midst of an intervention.

Manual registration accuracies were similar and averaged within 10 mm across the three manual registration methods and training levels using the HoloLens 2 system. The accuracies on HoloLens 2 are congruent to previously substantiated sub-centimeter accuracies reported with HoloLens 1 [19–24]. A trend of improving registration accuracies over consecutive attempts was suggested but not statistically significant. Of note, the accuracies reported were measured based on a 2D image and do not represent true 3D accuracies. Currently, a standard for measuring 3D accuracy for AR headset devices does not exist. Further work will involve the use of a surgical-grade optical tracking system (Polaris Vega, NDI, Waterloo, Canada) to assess true 3D accuracies. In addition, automatic registration accuracies were unable to be captured using HoloLens's built-in camera as the camera was actively being used for automated image detection and unavailable for image capture. Nevertheless, studies that have attempted to measure HoloLens accuracies in meaningful ways have shown accuracies and tracking improved with computer vision augmentation [21,33–35].

Registration data obtained from the two tertiary academic centers were analyzed individually. Registration times were similar for different methods of manual registration but not for automatic registration. Center 1's automatic registration times were more than twice of those obtained from Center 2. Examination of experiment protocols revealed that all testing at Center 1 were completed in dimly lit radiology reading rooms, optimized for image viewing on workstations while all testing at Center 2 were completed in brightly lit rooms. Previous work had demonstrated the effects of brightness on depth perception and accurate depth presentation [36]. Dim environments during automatic registration may have resulted in multiple re-aligning attempts by the participants. In addition, better contrast between the registration target and the background in brightly lit environments likely contributed to faster computer vision recognition. Furthermore, a HoloLens-based vascular localization system demonstrated larger precision errors when used without a surgical lamp compared with usage in the same operating room background brightness with the addition of a surgical lamp [37]. These effects should be taken into account for future AR implementation in different clinical settings. For example, in CT-guided and fluoroscopy-guided procedural suites, the operators could expect longer automatic registration times and potentially decreased registration accuracies given the low light environments. In operating rooms or outpatient office settings, the automatic registration times may be shorter and registration accuracies better.

There are limitations associated with this study. First, evaluation of HoloLens 2, similar to its previous generation, was performed on stationary objects with a fixated CT grid as the registration target. Registration time and accuracy with human participants, accounting for patient movement, respiratory motion, soft tissue deformation, and potential interference from procedural manipulation, were not assessed. Further testing on more realistic procedural settings, with developing technologies in motion compensation and deformable modeling, are needed. Second, the accuracy assessment was performed on a 2D plane rather than in the 3D space. Depth deviation could not be reliably assessed. Future testing with separate images taken from each monocular view may be obtained to calculate true registration alignment errors.

Multicenter assessment of HoloLens 2 for registering a 3D holographic model onto a CT grid showed that automatic registration through computer vision remains the optimal method, with performance dependence on background brightness. Manual registration times decreased significantly with updated hand gesture manipulation on HoloLens 2 versus HoloLens 1, with improvements over consecutive attempts approaching the registration times of automatic registration and outperforming external controller mediated registration significantly. Accurate and fast registration will be crucial for implementation of AR for procedural guidance during image-guided interventions.

**Figures**

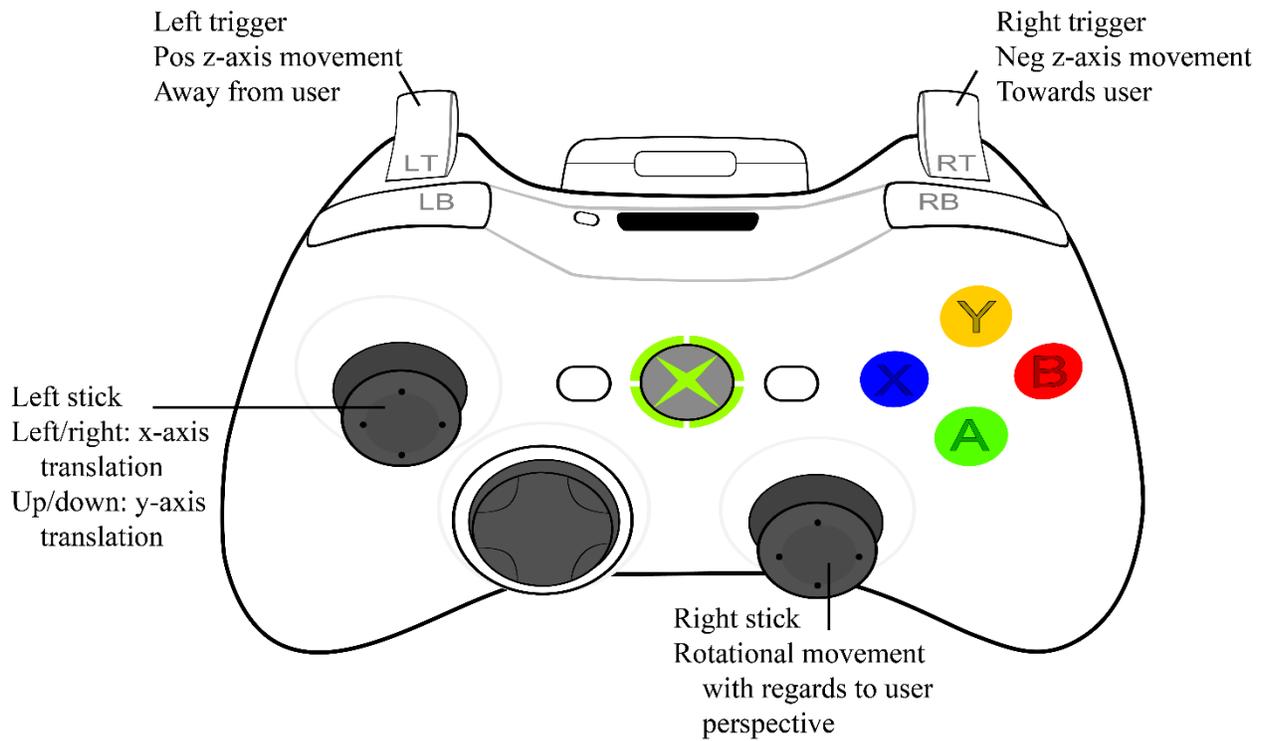

Figure 1. Three-dimensional hologram interactions mapped to Xbox gaming controller. Figure adapted from https://commons.wikimedia.org/wiki/File:360_controller.svg by Alphathon, CC BY-SA 3.0 <https://creativecommons.org/licenses/by-sa/3.0>, via Wikimedia Commons.

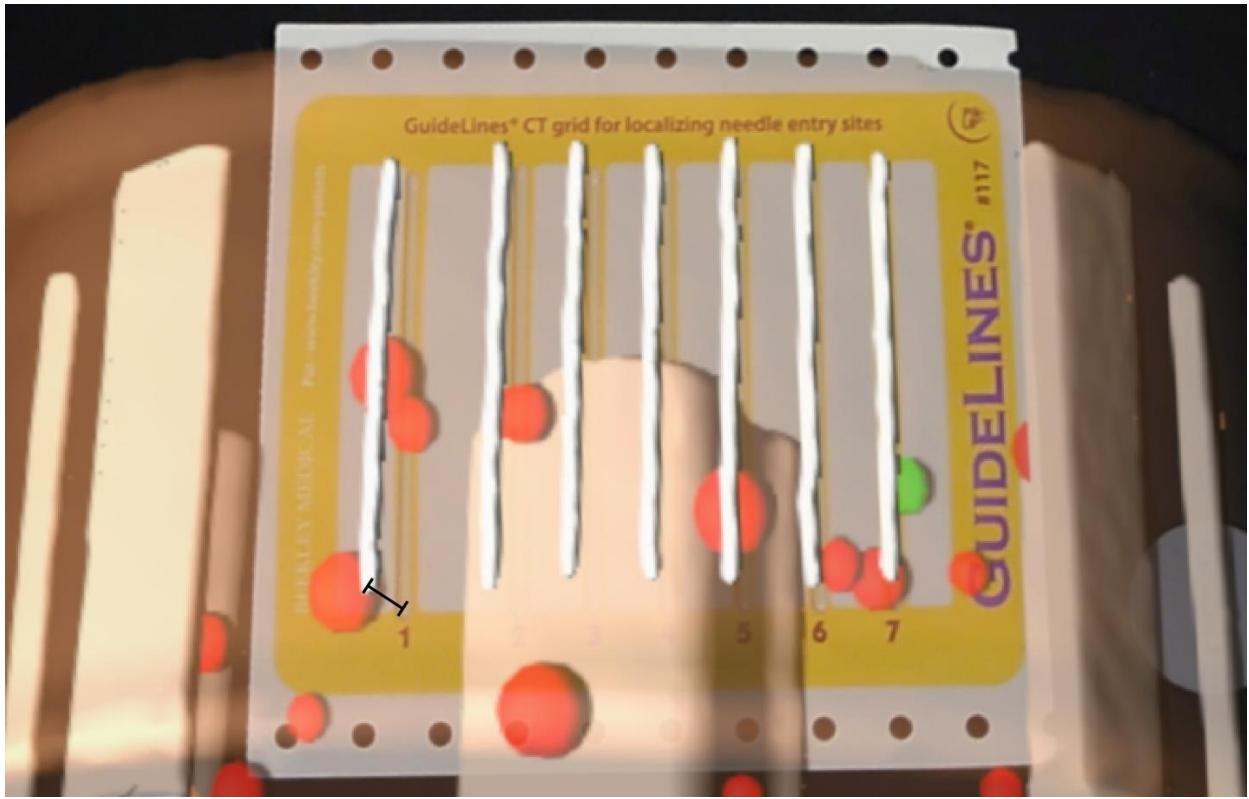

**Figure 2**. Example image capture displaying measurement of the registration error based on the imaged 2D plane.

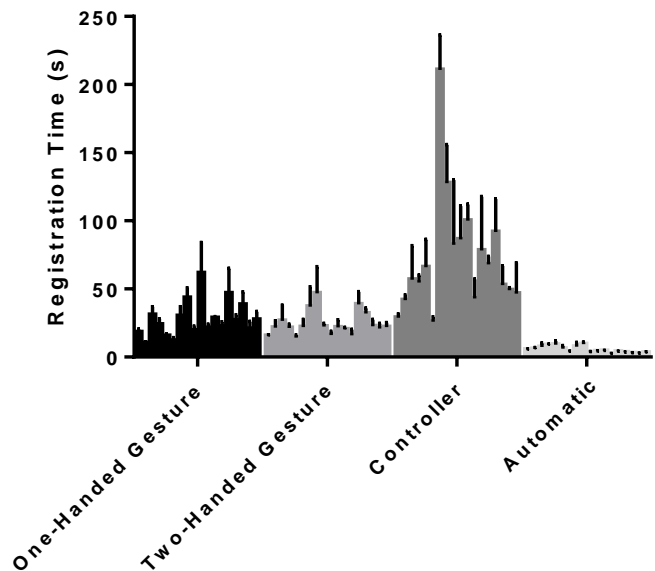

**Figure 3**. HoloLens 2 manual and automatic registration times from two centers. Each peak represented a single registration attempt.

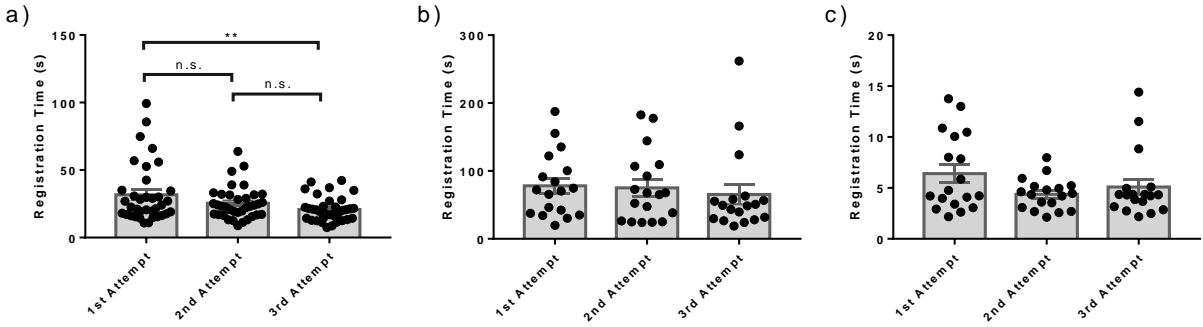

**Figure 4**. HoloLens 2 registration times by order of attempt. a) Hand gesture registration (p<0.01). b) Xbox controller registration (p>0.05). c) Automatic registration (p>0.05). n.s., non-significant. **, p<0.01.

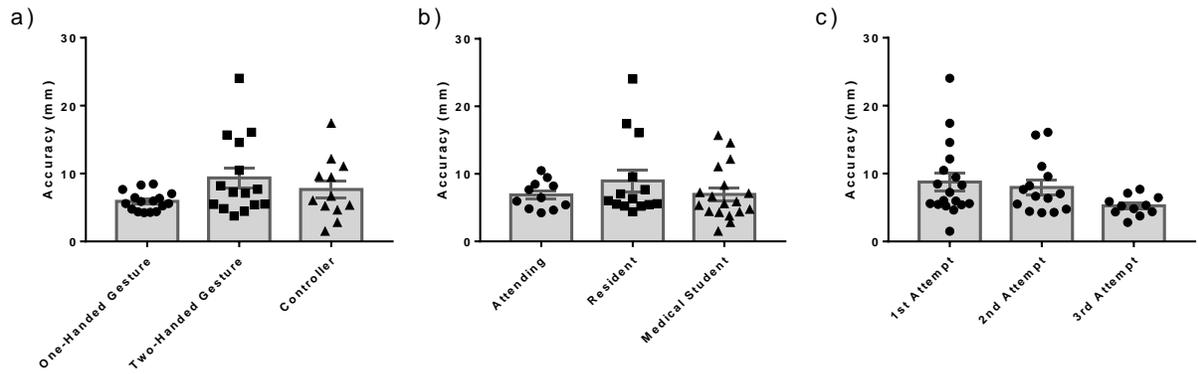

**Figure 5**. HoloLens 2 registration accuracy. a) Registration accuracy by registration methods (p>0.05). b) Registration accuracy by training level (p>0.05). c) Registration accuracy by number of attempts (p=0.05).

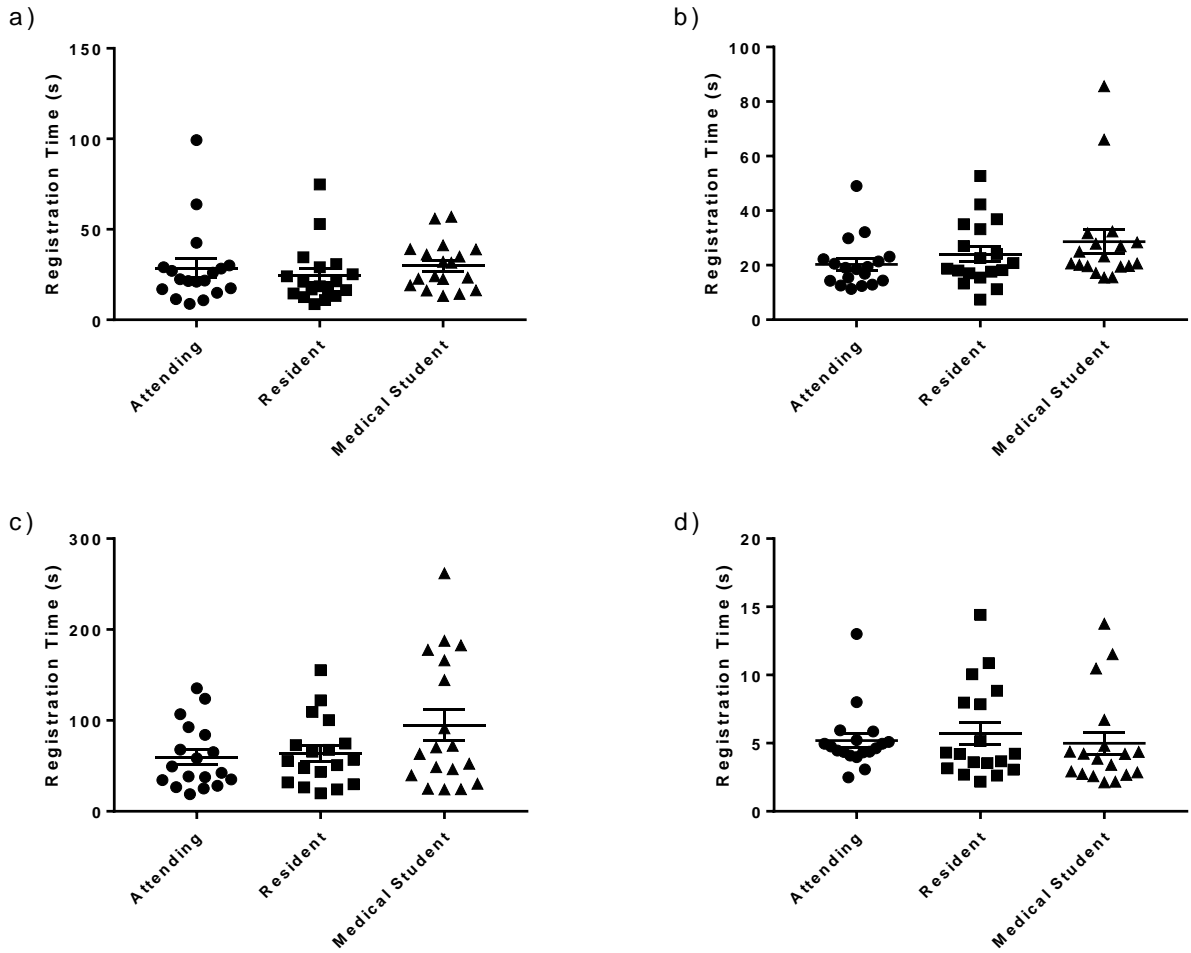

**Supplemental Figure1**. HoloLens 2 registration times by training level. a) One-handed gesture registration (p>0.05). b) Two-handed gesture registration (p>0.05). c) Xbox controller registration (p>0.05). d) Automatic registration (p>0.05).

**Table 1.** HoloLens 1 and HoloLens 2 Relevant Technical Specification Comparisons [25–30].

| Specifications | HoloLens 1 | HoloLens 2 |
| --- | --- | --- |
| Processor | Intel 32-bit (1 GHz) | Qualcomm Snapdragon 850 (3 GHz) |
| Field of View (Horizontal) | 34 ° | 52 ° |
| Field of View (Vertical) | 17.5 ° | 29 ° |
| Resolution (Pixels) | 1440 x 1440 | 2048 x 1080 |
| Refresh Rate | 90 Hz | 120 – 240 Hz |
| Battery Life | 2.5 Hours | 3 Hours |
| Cost | $3,000 | $3,500 |
| Release Year | 2016 | 2019 |
| Hand Tracking | Thumb and index finger recognition. Limited preset gesture recognition. Two-handed manipulation later introduced in 2018. | Two-handed fully articulated gesture tracking allowing up to 25 points of articulation per side including the wrist and fingers. |
| 3D Model Manipulation | Air tap and select. | Direct manipulation, including resizing, scaling, grabbing, pinching, dragging and holding |
| Tracking Degrees of Freedom | 6 | 6 |
| Interaction Range | Near-field only (nose to waist and between shoulders) | Near- and far-field interactions. |
| Eye Tracking | No | Yes |